\documentclass{article} %
\usepackage{iclr2024_conference,times}

\usepackage{amsmath,amsfonts,bm}

\def\eqref#1{equation~\ref{#1}}

\def\1{\bm{1}}

\DeclareMathAlphabet{\mathsfit}{\encodingdefault}{\sfdefault}{m}{sl}
\SetMathAlphabet{\mathsfit}{bold}{\encodingdefault}{\sfdefault}{bx}{n}

\newcommand{\defeq}{\coloneqq}

\usepackage{times}
\usepackage{latexsym}
\usepackage{tikz}
\usetikzlibrary{calc}
\usepackage{xspace}

\usepackage[small]{caption}
\usepackage{microtype}
\usepackage{graphicx}
\usepackage{subfigure}
\usepackage{booktabs} %
\usepackage[noend]{algpseudocode}
\usepackage{amsthm,amssymb,amsmath}
\usepackage{amsmath}
\usepackage{glossaries}
\usepackage{float}
\usepackage{listings}
\usepackage{hyperref}
\usepackage[capitalize]{cleveref}
\usepackage{url}
\usepackage{mdframed}
\usepackage{forest}
\usepackage{xspace}
\usepackage{siunitx}
\usepackage{inconsolata}
\usepackage{mathtools}
\usepackage{enumitem}
\usepackage{multirow}
\newtheorem{proposition}{Proposition}

\newcommand{\rowheight}{5.7mm}
\def\cboxwgt{9mm}
\newcommand{\cbox}[3]{%
    \definecolor{temp}{rgb}{#1}%
    \tikz[baseline,anchor=base] \node[fill=temp,text width=\cboxwgt,align=center,rounded corners=1.5pt,minimum height=\rowheight] (X) {\centering #2};%
}
\newcommand{\pmstack}[2]{\parbox[m]{8mm}{\centering#1\\[.2mm]\scriptsize$\pm$\,#2}}

\newcommand{\apjacob}[1]{\textcolor{teal}{}}
\newcommand{\jda}[1]{\textcolor{purple}{}}
\newcommand{\yk}[1]{\textcolor{green}{}}
\newcommand{\gabri}[1]{\textcolor{red}{}}

\title{The Consensus Game: Language Model \\ Generation via Equilibrium Search}

\author{Athul Paul Jacob\thanks{Correspondence to: apjacob@mit.edu} \\
MIT\\
\And
Yikang Shen \\
MIT-IBM AI Lab \\
\And 
Gabriele Farina \\
MIT \\
\And 
Jacob Andreas \\
MIT \\
}

\iclrfinalcopy %
\begin{document}

\colorlet{gencol}{teal!60!black}
\colorlet{discol}{brown!60!black}
\newcommand{\generator}{\textsc{\textcolor{gencol}{generator}}\xspace}
\newcommand{\discriminator}{\textsc{\textcolor{discol}{discriminator}}\xspace}
\newcommand{\vtrue}{\texttt{\textcolor{green!50!black}{correct}}\xspace}
\newcommand{\vfalse}{\texttt{\textcolor{purple!70!black}{incorrect}}\xspace}
\newcommand{\algo}{\textsc{equilibrium-ranking}\xspace}
\newcommand{\gen}{{\color{gencol}\textsf{G}}}
\newcommand{\dis}{{\color{discol}\textsf{D}}}
\newcommand{\correct}{\vtrue}
\newcommand{\incorrect}{\vfalse}

\newcommand{\truthobj}{\textcolor{black}{correctness parameter}\xspace}
\newcommand{\gamename}{\textsc{\textcolor{black}{consensus game}}\xspace}
\newcommand{\algog}{\textsc{equilibrium-ranking-generator}\xspace}
\newcommand{\algod}{\textsc{equilibrium-ranking-discriminator}\xspace}
\newcommand{\llamas}{LLaMA-7B\xspace}
\newcommand{\llamab}{LLaMA-13B\xspace}
\newcommand{\lm}{P_\textsf{LM}}
\newcommand{\query}{x}
\newcommand{\sequences}{\mathcal{Y}}
\maketitle

\begin{abstract}
    When applied to question answering and other text generation tasks, language models (LMs) may be queried \emph{generatively} (by sampling answers from their output distribution) or \emph{discriminatively} (by using them to score or rank a set of candidate outputs). These procedures sometimes yield very different predictions. How do we reconcile mutually incompatible scoring procedures to obtain coherent LM predictions?
    We introduce a new, a training-free, game-theoretic procedure for language model decoding. Our approach casts language model decoding as a regularized imperfect-information sequential signaling game---which we term the \gamename---in which a \generator seeks to communicate an abstract \truthobj using natural language sentences to a \discriminator. We develop computational procedures for finding approximate equilibria of this game, resulting in a decoding algorithm we call \algo.
    Applied to a large number of tasks (including reading comprehension, commonsense reasoning, mathematical problem-solving, and dialog), \algo consistently, and sometimes substantially, improves performance over existing LM decoding procedures---on multiple benchmarks, we observe that applying \algo to \llamas outperforms the much larger LLaMA-65B and PaLM-540B models.
    These results highlight the promise of game-theoretic tools for addressing fundamental challenges of truthfulness and consistency in LMs.
\end{abstract}

\section{Introduction}

Current language models (LMs) perform quite well on some tasks involving generation or verification of factual assertions---including question answering, fact-checking, and even unconditional text generation.
But they are far from perfectly reliable, and there is increasing evidence that LMs actually grow more prone to generating false but frequently repeated statements with increasing scale \citep{mckenzie2023inverse}.
Further complicating matters,
LMs offer multiple affordances for solving factual generation tasks. They may be used both \emph{generatively} (e.g.\ by asking for the most probable answer to a question) or \emph{discriminatively} (e.g.\ by presenting a (question, answer) pair and asking whether the answer is acceptable) and, these two procedures do not always produce consistent results: generative procedures may fail when probability mass is spread across multiple contradicting answers~\citep{wang2023selfconsistency,mitchell2022enhancing}, while discriminative procedures may fail due to miscalibration~\citep{han2022prototypical,chen2022close} or subtle dependence on question wording~\citep{jiang2020can}. Given these noisy and often-conflicting signals, how should we distill out an LM's best guess at the truth?

\begin{figure}[t]
    \centering
    \includegraphics[width=\columnwidth, trim=0.5cm 6cm 1.1cm 5.6cm,clip,scale=0.22]{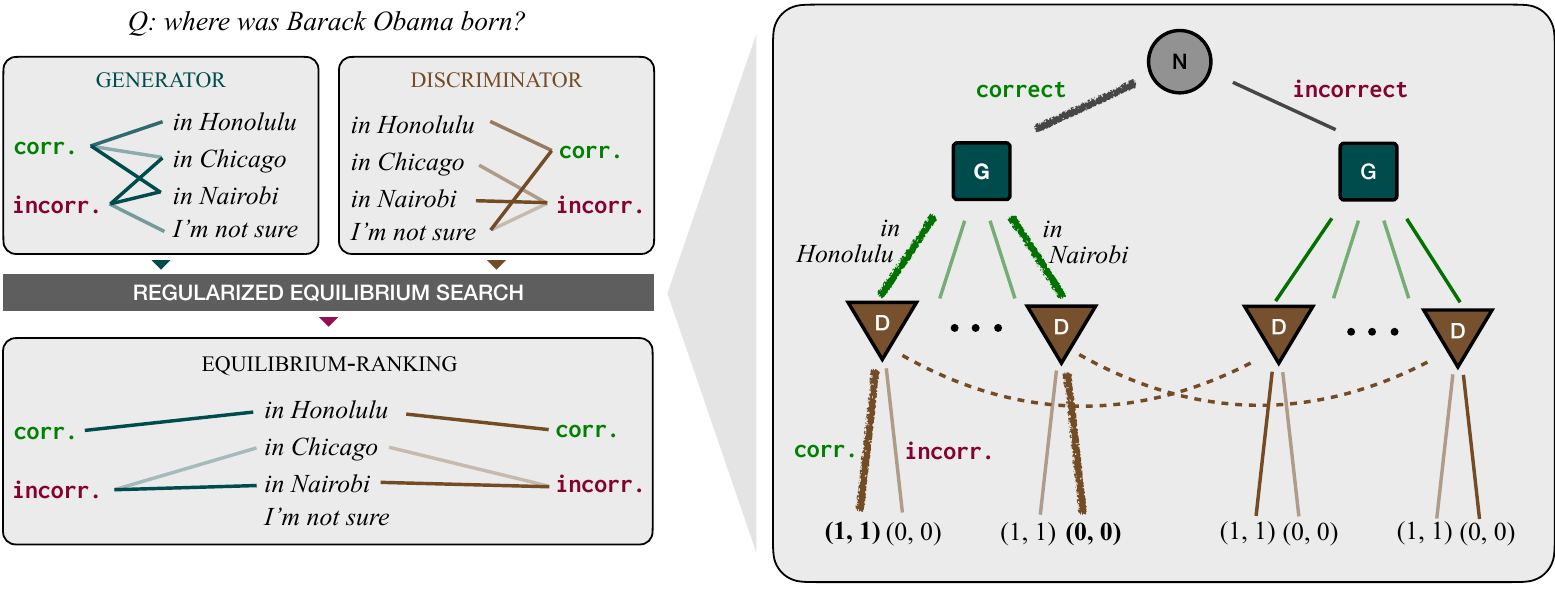}
    \caption{(Left) Overview of our approach. Differing LM queries fail to exhibit consensus about the answer to a factual question. By reconciling predictions between generative and discriminative LM queries using the \gamename, we obtain an accurate prediction. (Right) Structure of the \gamename, a two-player sequential signaling game with imperfect information. First, the environment (N) uniformly samples a correctness parameter. A \generator (G) conditioned on this parameter produces a natural language string from a set of candidates. The \discriminator (D) only observes this string and must predict the correctness parameter sampled by environment. If the \discriminator correctly identifies this parameter, then both players receive a reward of 1. The dashed line connects nodes that are indistinguishable by the \discriminator, since the \discriminator does not observe the \truthobj. By computing \emph{regularized equilibrium strategies} for this game, we obtain predictions that reflect a consensus between the \generator and \discriminator.}
    \label{fig:teaser}
\end{figure}

This paper presents an approach for reconciling generative and discriminative LM decoding procedures by formulating decoding as a signaling game \citep{lewis2008convention} that we call the \gamename. At a high level, this game features a \generator agent that must communicate an abstract \vtrue or \vfalse value to a \discriminator agent, but may only do so using a set of candidate natural language strings (\cref{fig:teaser}). Intuitively, an effective \emph{strategy} for this game (i.e.\ a joint policy) is one in which the \generator and \discriminator agree on the assignment of strings to correctness values. Given such a strategy, we may inspect it to identify candidates agreed by consensus to be \vtrue{}.

Doing so requires solving a multi-step game with a complex (string-valued) action space. In recent years,
\emph{no-regret learning} algorithms have emerged as the preferred technique to compute effective strategies for such games, and have been successfully deployed in Poker \citep{brown2018superhuman,brown2019superhuman}, Stratego \citep{perolat2022mastering}, and Diplomacy \citep{bakhtin2022mastering,meta2022human,Jacob22:Modeling}. Here, we show that they can also be applied to free-form language generation tasks.
We call this game-theoretic approach to LM decoding \algo. Applied in 6 question answering benchmarks: MMLU \citep{hendrycks2020measuring}, ARC \citep{clark2018think}, RACE \citep{lai2017race}, HHH \citep{askell2021general}, TruthfulQA \citep{lin2021truthfulqa} and, GSM8K~\citep{cobbe2021training}, \algo offers substantial improvements over existing generative, discriminative, and mixed decoding procedures.
More generally, our results highlight the usefulness of the game-theoretic toolkit for formalizing and improving coherence in LMs. Improved coherence in turn leads to improved accuracy on factual tasks.

\section{Language Model Consensus as Equilibrium Search}
\label{sec:formulation}

We study the problem of obtaining correct output from a \textbf{language model}, which maps input strings $x$ to output strings $y$ according to some distribution $\lm(y \mid x)$. While the techniques we present here are general, we focus in this paper on \textbf{question answering} problems consisting of a query $\query$ (\emph{In which of the following cities was Barack Obama born?}) and a set of candidate answers $\sequences$ (\emph{Honolulu, Chicago, \ldots}) which may themselves have been sampled from the complete $\lm(\cdot \mid \query)$. Given a set of candidates, we may them with an LM in (at least) two ways:

\begin{itemize}[left=5mm]
    \item \emph{Generatively}, by supplying as input (i)
          the query $\query$, (ii) the set of candidates $\sequences$, and (iii) a natural language prompt indicating that a correct answer is desired. In this case, the LM may be thought of as modeling a distribution $\lm(y \mid \query, \vtrue)$, where the token $\vtrue$ denotes the fact that the model was prompted to generate a correct answer.
    \item \emph{Discriminatively}, by supplying as input (i) the query $\query$ and (ii) a possible candidate answer $y \in \sequences$, together with (iii) a prompt indicating that a correctness assessment $v \in \{\vtrue,\vfalse\}$ is sought. In this case, the language model acts as a
          model of as modeling a distribution
          $\lm(v \mid \query, y)$ where $v \in \{\vtrue, \vfalse\}$.
\end{itemize}

These two approaches are conceptually equivalent. But as noted in the introduction, current LMs may give very different answers when queried in these different ways:
answers produced generatively might be assessed \incorrect with high probability or vice-versa. %
Research on LMs has proposed two broad solutions to this problem. \textbf{Ensembling methods} \citep{ouyang2022training, li2016mutual, glaese2022improving} simply combine discriminative and generative scores directly. While moderately effective, such approaches suffer from the fact that LM predictions are often poorly calibrated both within and across contexts, meaning that scores may not combine in meaningful or consistent ways. \textbf{Deliberation methods} \citep{wei2022chain,yao2023tree,du2023improving} perform this reconciliation within the LM itself, e.g.\ by re-prompting with competing inputs and an instruction to generate a textual justification for the best one. Such methods incur significant computational overhead.\footnote{As shown in \cref{sec:gsm}, they are also orthogonal to, and composable with, the approach we propose here.}

How might we design a principled and computationally efficient procedure for obtaining a ``consensus'' between competing LM predictions? Informally, a consensus prediction would satisfy two key properties: \textbf{coherence} (generative and discriminative scoring procedures should agree about which candidate answers are correct) and \textbf{reasonableness} (predictions should not be arbitrary, but instead as close as possible to original generator / discriminator behavior).
The key idea in this paper is to
operationalize these high-level desiderata in language of game theory, using \textbf{regularized equilibrium} concepts as formal framework for defining both coherence and reasonableness. Below, we introduce and explain this framework in detail,
describing how to instantiate decoding as a signaling game, then compute equilibrium strategies of this game to obtain consensus LM predictions.

\subsection{The \gamename}

Our approach to language generation begins by formulating language generation as a signaling game \citep{lewis2008convention} that we call the \gamename.
The \gamename is played on a game tree, as depicted in Figure~\ref{fig:teaser}.
At the start of the game (that is, at the root of the game tree), a \emph{\truthobj} $v \in \{\correct,\incorrect\}$ is selected uniformly at random by the environment. The \truthobj is observed only by the \generator, and controls whether the \generator should aim to generate \emph{\correct} or \emph{\incorrect} answers. Upon observing this parameter, the \generator produces a natural language string from a fixed set of candidates. Finally, this string is observed by the \discriminator, who tries to guess the value of the \truthobj by selecting one of $\{\vtrue, \vfalse\}$ as an answer.
Both players obtain a \textbf{payoff} of 1 if the \discriminator correctly identifies the value of the \truthobj, 0 otherwise.

With this definition, it may be observed that players' expected \textbf{utilities} (the payoffs they may expect to receive) are as follows:
\begin{align*}
    u_\gen(\pi_\gen, \pi_\dis) & \defeq \frac{1}{2}\sum_{v \in \{\vtrue, \vfalse\}}\sum_{y\in \sequences} \pi_\gen(y \mid v) \cdot \pi_\dis(v \mid y) , \\
    u_\dis(\pi_\gen, \pi_\dis) & \defeq \frac{1}{2}\sum_{v \in \{\vtrue, \vfalse\}}\sum_{y\in \sequences} \pi_\gen(y \mid v) \cdot \pi_\dis(v \mid y).
\end{align*}

What is an effective strategy for maximizing these utilities?
A standard answer to this question in the game theory literature is that a \textbf{Nash equilibrium} of the game should be sought. A Nash equilibrium is a pair of policies---one for the \generator and one for the \discriminator---such that each policy is optimal. That is, each player's strategy maximizes their expected given the other player's strategy.
At a Nash equilibrium, no player has an incentive to unilaterally behave in any other way. In signaling games, Nash equilibria offer a natural way of formalizing the \textbf{coherence} criterion above: at equilibrium, both the \generator and \discriminator must agree about which messages correspond to \vtrue and \vfalse respectively in order to obtain a nonzero payoff.

However, Nash equilbria of the \gamename are not guaranteed to provide the second criterion of \textbf{reasonableness}. This is because  the \gamename admits a multitude of Nash equilibria that are incompatible with the common-sense notion of truthfulness. For example, the strategy in which the \generator deterministically maps \correct $\mapsto$ ``Nairobi'', \incorrect $\mapsto$ ``Honolulu'', and the \discriminator maps ``Nairobi'' $\mapsto$ \correct, ``Honolulu'' $\mapsto$ \incorrect forms a Nash equilibrium.

In order to sidestep the inappropriate equilibria and ensure reasonableness, we introduce a \textbf{regularization term} in the utility of the players, so that both the \generator and the \discriminator are penalized for settling on strategies that are far from some pair of \emph{initial policies}: $\smash{\pi^{(1)}_\gen}$ and $\smash{\pi^{(1)}_\dis}$. By parameterizing these policies using a pre-trained LM, we may use knowledge about what answers are likely to be correct \emph{a priori} to guide selection of an equilibrium.
As in \citet{Jacob22:Modeling}, we incorporate this regularization term directly into the utility function (payoff) that the \generator and \discriminator attempt to optimize. Rather than the simple 0--1 payoff determined by agreement on the correctness parameter, they now attempt to optimize:
\begin{align*}
    u_\gen(\pi_\gen, \pi_\dis) & \defeq -\lambda_\gen \cdot \mathrm{D}_\mathrm{KL}[\pi_\gen(\cdot \mid v), \pi^{(1)}_\gen(\cdot \mid x, v)] + \frac{1}{2}\sum_{v}\sum_{y\in \sequences} \pi_\gen(y \mid \query, v) \cdot \pi_\dis(v \mid \query, y) , \\
    u_\dis(\pi_\gen, \pi_\dis) & \defeq -\lambda_\dis \cdot \mathrm{D}_\mathrm{KL}[\pi_\dis(\cdot \mid y), \pi^{(1)}_\dis(\cdot \mid x, y)] + \frac{1}{2}\sum_{v}\sum_{y\in \sequences} \pi_\gen(y \mid \query, v) \cdot \pi_\dis(v \mid \query, y).
\end{align*}
Note that the initial policies $\smash{\pi^{(1)}_\gen}(y \mid x, v)$
and $\smash{\pi^{(1)}_L(v \mid x, y)}$ may be derived from an LM \emph{prompted} with some initial string $x$, in order to obtain context-predictions (e.g.\ answers to a question).
With these utilities, Nash equilibria for the game are pulled by the initial \generator and \discriminator policies in the direction of increased consensus.

\citet{bakhtin2022mastering} and \citet{meta2022human} employed a similar regularization method for choosing actions, rather than messages, in versions of the board game Diplomacy. \citet{franke2013game, franke2017game} have explored signaling games in the context of linguistic pragmatics to explain human language use. To the best of our knowledge, however, this is the first proposal for using regularized equilibrium concepts in signaling games to define target behavior in a language generation task.

\subsection{\algo: LM Ranking via equilibrium search}
With this formulation, text generation requires finding a Nash equilibrium of the game with the utilities given above. How should we compute such an equilibrium?
No-regret learning algorithms have emerged in recent years as the preferred technique to approximate equilibria in large games, and have been successfully
employed to solve games at human or even superhuman level.
At a high level, these algorithms find equilibrium by repeatedly interacting in the game and refining their policies after each iteration $t$.
so as to minimize \textbf{regret} (the gap between the chosen action and the best action in hindsight). %

In this section, we describe in detail how to perform no-regret learning in the \gamename in order to obtain consensus policies.
Importantly, this approach modifies only signalling policies, and not the base policies $\smash{\pi^{(1)}_\gen}$ and $\smash{\pi^{(1)}_\dis}$ (i.e.\ the LM).
In this sense, generating text by performing no-regret learning in the \gamename might be described as a \emph{training-free consensus-planning method}. We call this method \algo.

\paragraph{Initial policies}
\label{subsec:speaker}
At time $t=1$, that is, before any equilibrium computation has happened, \algo defines the initial policies $\pi^{(1)}_\gen$ and $\pi^{(1)}_\dis$ of the \generator and \discriminator, respectively, as follows. $\pi^{(1)}_\gen$ normalizes $\lm$\footnote{In ARC, RACE, HHH, TruthfulQA, and GSM8K, based on prior work \citep{touvron2023llama, brown2020language}, we additionally normalize $\lm(u|\query)$ by the likelihood of the completion given “Answer:” as context: $\lm(u \mid \text{"Answer:"})$.} across $v$ and $y$:
\begin{align*}
    \pi^{(1)}_\gen (y \mid \query, v) \propto \frac{\lm(y \mid \query, v)}{\sum_{v'} \lm(y \mid \query, v')}.
\end{align*}

Similarly for the \discriminator, the initial policy normalizes across $y$ and $v$:
\begin{align*}
    \pi^{(1)}_\dis (v \mid \query,y) \propto \frac{\lm(v \mid \query,y)}{\sum_{y'} \lm(v \mid \query,y')}.
\end{align*}
This crucial step enables us to extract a well calibrated \generator and \discriminator from $\lm$. The specific form of the \generator incorporates $v = \vfalse$, and this resembles contrastive decoding \citep{li2022contrastive}, where they rely on a weaker LM as opposed to an LM conditioned on $\vfalse$ (See, \cref{sec:results} for details). This \discriminator resembles approaches that query the LM itself to produce critiques \citep{ganguli2023capacity, chen2023teaching, yao2023tree}. However, to the best of our knowledge, this specific instantiation has not been explored in the past.

\paragraph{Evolution of policies}
A classic observation in the theory of imperfect-information sequential games is that minimization of regret (viewed as a function of their overall policy on the game tree) can be achieved by solving separate, \emph{local}, regret minimization problems at each information set (\textit{i.e.}, decision point) of the game. This observation underpins the CFR framework \citep{Zinkevich07:Regret}, as well as its generalization to more general convex losses, known as laminar regret decomposition \citep{Farina19:Online}. In our case, these techniques enable us to decompose the policy update of the players into separate updates for each \truthobj $v$ (for the \generator) and for each sequence $y$ (for the \discriminator). We provide more detail and background in Appendix~\ref{app:cfr}.

In our setting, after operating the regret decomposition step, we find that the local regret minimization problems are composed of a bilinear term, plus a strongly convex KL-regularization term. Such composite utilities can be handled by the piKL algorithm \citep{Jacob22:Modeling}, which is specifically designed to perform regret minimization on KL-regularized objectives.
In our setting, piKL prescribes that each player keep track of their average values:
\begin{align*}
    Q^{(t)}_\gen(y \mid \query, v) \defeq \frac{1}{2t}\sum_{\tau = 1}^t \pi^{(\tau)}_\dis(v \mid \query, y), \qquad
    Q^{(t)}_\dis(v \mid \query, y) \defeq \frac{1}{2t}\sum_{\tau = 1}^t \pi^{(\tau)}_\gen(y \mid \query, v).
\end{align*}
Each player then updates their policy according to:
\begin{align}
    \pi_\gen^{(t+1)}(y \mid \query, v) & \propto \exp\left\{\frac{Q^{(t)}_\gen(y \mid \query, v) + \lambda_\gen \log \pi_\gen^{(1)}(y \mid \query, v)}{1/(\eta_\gen t) + \lambda_\gen} \right\}, \label{eq:update G} \\
    \pi_\dis^{(t+1)}(v \mid \query, y) & \propto \exp\left\{\frac{Q^{(t)}_\dis(v \mid \query, y) + \lambda_\dis \log \pi_\dis^{(1)}(v \mid \query, y)}{1/(\eta_\dis t) + \lambda_\dis} \right\}, \label{eq:update D}
\end{align}
where $\eta_\gen,\eta_\dis > 0$ are \emph{learning rate} hyperparameters. piKL no-regret dynamics are known to have strong guarantees, including the following (more formal statements about the guarantees are available in Appendix~\ref{app:cfr}):
\begin{itemize}[nosep,left=5mm]
    \item \textbf{Convergence to an equilibrium point}. The average correlated distribution of play of \generator and \discriminator converges to the set of (regularized) coarse-correlated equilibria of the game.
    \item \textbf{Regularization toward reasonableness}. The average policy of any player remains within a radius of size roughly $1/\lambda_i$ from the initial policy $\pi_i^{(1)}$, where $\lambda_i$ is the amount of regularization applied to any player $i\in\{\generator,\discriminator\}$ (see \cref{prop:radius}).
    \item \textbf{Avoidance of regret}. The cumulative regret incurred by each of the players grows only logarithmic in the number of training steps (see \cref{prop:regret}).
\end{itemize}
At convergence, \algo returns $\pi_\gen^{*}$ and $\pi_\dis^{*}$, which are the refined \generator and \discriminator. We also remark that a recent result by \citet{Anagnostides22:Last-Iterate} showed that as long as the regularization function has Lipschitz-continuous gradients (a condition that can be easily achieved by introducing a small perturbation in the KL regularization term), decentralized learning dynamics similar to piKL converge in iterates to a regularized Nash equilibrium of the \gamename, since it is a potential game~\citep{Monderer96:Potential}. In practice, we do not investigate introducing a small perturbation, as we witness good convergence properties in practice even without any perturbation. As mentioned earlier, convergence to a regularized Nash equilibrium is important to guarantee both coherence and reasonableness. Extensive empirical validation presented in the next section demonstrates the benefits of this approach in practice. \gabri{Explicitly mention that we leave the question of proving last-iterate convergence in our setting open?}

\paragraph{Computational cost of our method.} At each iteration, our method needs to update the policies $Q^{(t)}_\gen, Q^{(t)}_\dis$ according to~(\ref{eq:update G}) and ~(\ref{eq:update D}). The number of operations at each iteration of the method is therefore linear in the number $|\sequences|$ of sequences available to the \generator.

\section{Experiments}
\label{sec:results}
As discussed in the previous section, \algo focuses on improving the \emph{correctness} of language models in question-answering (QA) tasks. However, correctness manifests in various forms across different domains, including truthfulness, factuality, valid reasoning, value alignment, among others. Therefore, we will evaluate its performance on a diverse set of QA tasks: MMLU \citep{hendrycks2020measuring}, ARC \citep{clark2018think}, RACE \citep{lai2017race}, HHH \citep{askell2021general}, and TruthfulQA \citep{lin2021truthfulqa}. It's important to note that \algo is a sampling strategy and not a delibration method like chain-of-thought (CoT) \citep{wei2022chain} and self-consistency \citep{wang2023selfconsistency}. Nevertheless, we will demonstrate in GSM8K \citep{cobbe2021training} that we can achieve some additional gains when combining \algo with self-consistency and CoT.

\setlength{\tabcolsep}{1.2mm}
\paragraph{Hyperparameters} \algo has four parameters, $\eta_\dis,\lambda_\dis$ and $\eta_\gen,\lambda_\gen$. Although tuning these parameters will lead to better performance, in all our experiments we set $\eta_\dis = \lambda_\dis = \eta_\gen = \lambda_\gen = 0.1$. We run \algo for 5000 iterations \footnote{ As remarked at the end of the previous section, each iteration of the learning process requires a number of floating-point operations that is linear in the number $|\sequences|$ available to the \generator. In most of our settings, $|\sequences| = 4$, making the overhead from  the learning dynamics on the \gamename negligible compared to the cost of inference for the language model. As such, even with an unoptimized implementation of the dynamics (\ref{eq:update G},\ref{eq:update D}), we observe that the computational cost associated with each iteration of the learning process takes about 40 microseconds on average.} %
\paragraph{Actions in the \gamename} As mentioned in Section~\ref{sec:formulation}, in order to make our approach amenable to current computational techniques, we make the modeling assumption that the \generator picks distribution over a finite set of candidates $\sequences$. In multiple-choices tasks, these are the multiple choice options. In generative tasks, a common approach to generate the finite set of candidates is via sampling with nucleus \citep{nucleus} and top-$k$ \citep{topk} from the distribution $\lm(y \mid q, \correct)$ where $y \in \sequences$. This is exactly the approach we use in our experiments, with $p=0.9$ for nucleus sampling and $k=50$.
\paragraph{Models} We use the 7B and 13B parameter models from the LLaMA family \citep{touvron2023llama} and perform 16-bit inference for all our experiments.
\paragraph{Prompting for \correct and \incorrect answers}
In our work, unless otherwise specified, conditioning on $(\query,\vtrue)$ for the $\lm$ corresponds to the standard zero-shot prompt. Similarly, conditioning on $(\query, \vfalse)$ is similar to $(\query, \vtrue)$ with the only difference that \textit{"Answer:"} is replaced with \textit{"Incorrect Answer:"} in the prompt.
\paragraph{Baselines}
In the multiple-choice based datasets (ARC, RACE, HHH, MMLU), we consider the following approaches:
\begin{itemize}[left=5mm]
    \item \textbf{Generative Ranking (G):} This baseline \citep{brown2020language,touvron2023llama} ranks every candidate $y$ by $P_{\text{LM}}(y \mid \query, \vtrue)$ and picks the top candidate. This is the standard approach used in past work. Due to implementational differences, when available, we include both official scores and our version.
    \item \textbf{Mutual Information Ranking (MI):} This mutual-information based \citep{li2016mutual} baseline is an ensemble-based approach that reweights every candidate $y$ by $P_{\text{LM}}(y \mid \query, \vtrue)\cdot P_{\text{LM}}(\vtrue \mid \query, y)$.
    \item \textbf{Self-Contrastive Ranking (SC):} This approach utilizes the normalized generator $\pi^{(1)}_\gen$ to reweight every candidate $y$ by $\smash{\pi^{(1)}_\gen (\vtrue \mid \query, y)}$. As discussed in \cref{sec:formulation}, this shares similarities with contrastive decoding~\citep{li2022contrastive}.
    \item \textbf{Discriminative Ranking (D):} This approach reweights every query-candidate pair $(\query, y)$ by $\pi^{(1)}_\dis (\vtrue \mid \query, y)$.
    \item \textbf{Equilibrium Ranking Generator (ER-G):} Similar to \textbf{SC}, this approach utilizes the final \algo-based generator $\pi^*_\gen$ to reweight every candidate $y$ by $\pi^*_\gen (y \mid \query, \vtrue)$.
    \item \textbf{Equilibrium Ranking Discriminator (ER-D):} Similar to \textbf{D}, this approach utilizes the final \algo-based discriminator $\pi^*_\dis$. This approach reweights every query-candidate pair $(\query, y)$ by $\pi^*_\dis (\vtrue \mid \query, y)$.
\end{itemize}

In free-form text generation tasks (TruthfulQA, GSM8K), we additionally consider \textbf{greedy decoding}. In the mathematical reasoning task (GSM8K), we also consider \textbf{self-consistency} \citep{wang2023selfconsistency}.

\begin{table*}[h]
    \def\cboxwgt{9mm}
\def\sp{1.4mm}
\begin{tabular}{llccccc|cc}
    \toprule
                                                                   &           &                                         &                                                                            &                                                                                            &                                                                                            &                                                                                             & \multicolumn{2}{c}{\textbf{Equil. ranking}}                                                                                                                                                                 \\
    Domain                                                         & Model     & G$^*$                                   & G                                                                          & MI                                                                                         & SC                                                                                         & D                                                                                           & ER-G                                                                                                 & ER-D                                                                                                 \\
    \midrule
    \multirow{2}{*}[-1mm]{\parbox[m][][c]{1.8cm}{\centering MMLU}} & LLaMA-7B  &             --                            & \cbox{0.9388081507112649,0.9408381391772395,0.9390849673202615}{30.4}{0.0} & \cbox{0.777762399077278,0.9178316032295271,0.7968627450980392}{33.1}{2.700000000000003}    & \cbox{0.9340715109573241,0.9401614763552479,0.9349019607843138}{30.5}{0.10000000000000142} & \cbox{0.33725490196078434,0.8549019607843137,0.40784313725490196}{\textbf{40.4}}{10.0}      & \cbox{0.3940945790080739,0.8630219146482122,0.4580392156862745}{39.4}{9.0}                           & \cbox{0.36567474048442916,0.858961937716263,0.4329411764705883}{39.9}{9.5}                           \\[\sp] %
                                                                   & LLaMA-13B &               --                          & \cbox{0.9388081507112649,0.9408381391772395,0.9390849673202615}{41.7}{0.0} & \cbox{0.9340715109573241,0.9401614763552479,0.9349019607843138}{41.8}{0.09999999999999432} & \cbox{0.9388081507112649,0.9408381391772395,0.9390849673202615}{41.7}{0.0}                 & \cbox{0.9293348712033833,0.9394848135332564,0.930718954248366}{41.9}{0.19999999999999574}   & \cbox{0.7493425605536332,0.9137716262975778,0.7717647058823529}{44.9}{3.1999999999999957}            & \cbox{0.7351326412918109,0.9117416378316032,0.7592156862745097}{\textbf{45.1}}{3.3999999999999986}   \\
    \midrule
    \multirow{2}{*}[-1mm]{\parbox[m][][c]{1.8cm}{\centering ARC                                                                                                                                                                                                                                                                                                                                                                                                                                                                                                                                                                                                                                             \\[\sp] Easy}}      & LLaMA-7B  & \cbox{1,1,1}{72.8}{4.599999999999994}   & \cbox{0.9388081507112649,0.9408381391772395,0.9390849673202615}{68.2}{0.0} & \cbox{0.9056516724336794,0.9361014994232987,0.9098039215686274}{68.8}{0.5999999999999943}   & \cbox{0.8630219146482122,0.9300115340253748,0.872156862745098}{69.5}{1.2999999999999972}   & \cbox{1.0,0.6196078431372549,0.6196078431372549}{52.5}{-15.700000000000003}                 & \cbox{0.7351326412918109,0.9117416378316032,0.7592156862745097}{\textbf{71.6}}{3.3999999999999915}   & \cbox{0.7398692810457517,0.9124183006535947,0.7633986928104575}{71.5}{3.299999999999997}             \\[\sp]
                                                                   & LLaMA-13B & \cbox{1,1,1}{74.8}{3.5999999999999943}  & \cbox{0.9388081507112649,0.9408381391772395,0.9390849673202615}{71.2}{0.0} & \cbox{0.9245982314494425,0.9388081507112649,0.9265359477124183}{71.5}{0.29999999999999716} & \cbox{0.8298654363706267,0.9252748942714341,0.842875816993464}{73.0}{1.7999999999999972}   & \cbox{0.9778546712802768,0.7406689734717417,0.7406689734717417}{65.0}{-6.200000000000003}   & \cbox{0.6451364859669357,0.8988850442137639,0.6797385620915033}{76.1}{4.8999999999999915}            & \cbox{0.6261899269511726,0.8961783929257977,0.6630065359477124}{\textbf{76.4}}{5.200000000000003}    \\
    \midrule
    \multirow{2}{*}[-1mm]{\parbox[m][][c]{1.8cm}{\centering ARC                                                                                                                                                                                                                                                                                                                                                                                                                                                                                                                                                                                                                                             \\[\sp] Challenge}} & LLaMA-7B  & \cbox{1,1,1}{47.6}{0.30000000000000426} & \cbox{0.9388081507112649,0.9408381391772395,0.9390849673202615}{47.3}{0.0} & \cbox{0.9340715109573241,0.9401614763552479,0.9349019607843138}{47.4}{0.10000000000000142}  & \cbox{0.3846212995001923,0.8616685890042292,0.4496732026143791}{56.5}{9.200000000000003}   & \cbox{0.9681660899653979,0.7936332179930796,0.7936332179930796}{42.7}{-4.599999999999994}   & \cbox{0.33725490196078434,0.8549019607843137,0.40784313725490196}{\textbf{58.7}}{11.400000000000006} & \cbox{0.33725490196078434,0.8549019607843137,0.40784313725490196}{58.3}{11.0}                        \\[\sp]
                                                                   & LLaMA-13B & \cbox{1,1,1}{52.7}{0.8000000000000043}  & \cbox{0.9388081507112649,0.9408381391772395,0.9390849673202615}{51.9}{0.0} & \cbox{0.9293348712033833,0.9394848135332564,0.930718954248366}{52.1}{0.20000000000000284}  & \cbox{0.49356401384083043,0.8772318339100346,0.5458823529411765}{59.3}{7.399999999999999}  & \cbox{0.9612456747404844,0.8314648212226067,0.8314648212226067}{48.5}{-3.3999999999999986}  & \cbox{0.3846212995001923,0.8616685890042292,0.4496732026143791}{61.1}{9.200000000000003}             & \cbox{0.36567474048442916,0.858961937716263,0.4329411764705883}{\textbf{61.4}}{9.5}                  \\
    \midrule
    \multirow{2}{*}[-1mm]{\parbox[m][][c]{1.8cm}{\centering RACE                                                                                                                                                                                                                                                                                                                                                                                                                                                                                                                                                                                                                                            \\[\sp] Middle}}   & LLaMA-7B  & \cbox{1,1,1}{61.1}{3.3999999999999986}  & \cbox{0.9388081507112649,0.9408381391772395,0.9390849673202615}{57.7}{0.0} & \cbox{0.9388081507112649,0.9408381391772395,0.9390849673202615}{57.7}{0.0}                  & \cbox{0.777762399077278,0.9178316032295271,0.7968627450980392}{60.4}{2.6999999999999957}   & \cbox{0.9778546712802768,0.7406689734717417,0.7406689734717417}{51.5}{-6.200000000000003}   & \cbox{0.6072433679354095,0.8934717416378316,0.6462745098039215}{63.2}{5.5}                           & \cbox{0.5882968089196463,0.8907650903498654,0.6295424836601307}{\textbf{63.5}}{5.799999999999997}    \\[\sp]
                                                                   & LLaMA-13B & \cbox{1,1,1}{61.6}{1.5}                 & \cbox{0.9388081507112649,0.9408381391772395,0.9390849673202615}{60.1}{0.0} & \cbox{0.9340715109573241,0.9401614763552479,0.9349019607843138}{60.2}{0.10000000000000142} & \cbox{0.6546097654748174,0.900238369857747,0.6881045751633987}{64.8}{4.699999999999996}    & \cbox{0.9520184544405997,0.8819069588619761,0.8819069588619761}{58.3}{-1.8000000000000043}  & \cbox{0.4698808150711265,0.8738485198000768,0.5249673202614379}{67.9}{7.800000000000004}             & \cbox{0.4272510572856594,0.867758554402153,0.4873202614379085}{\textbf{68.6}}{8.499999999999993}     \\
    \midrule
    \multirow{2}{*}[-1mm]{\parbox[m][][c]{1.8cm}{\centering RACE                                                                                                                                                                                                                                                                                                                                                                                                                                                                                                                                                                                                                                            \\[\sp] High}}     & LLaMA-7B  & \cbox{1,1,1}{46.9}{0.5}                 & \cbox{0.9388081507112649,0.9408381391772395,0.9390849673202615}{46.4}{0.0} & \cbox{0.9418685121107266,0.9373933102652826,0.9373933102652826}{46.3}{-0.10000000000000142} & \cbox{0.5361937716262977,0.8833217993079584,0.5835294117647059}{53.1}{6.700000000000003}   & \cbox{0.9437139561707035,0.9273048827374086,0.9273048827374086}{46.0}{-0.3999999999999986}  & \cbox{0.3419915417147251,0.8555786236063052,0.4120261437908497}{56.3}{9.899999999999999}             & \cbox{0.33725490196078434,0.8549019607843137,0.40784313725490196}{\textbf{56.4}}{10.0}               \\[\sp]
                                                                   & LLaMA-13B & \cbox{1,1,1}{47.2}{-0.6999999999999957} & \cbox{0.9388081507112649,0.9408381391772395,0.9390849673202615}{47.9}{0.0} & \cbox{0.9103883121876202,0.9367781622452903,0.9139869281045752}{48.4}{0.5}                 & \cbox{0.33725490196078434,0.8549019607843137,0.40784313725490196}{58.9}{11.0}              & \cbox{0.5030372933487121,0.8785851595540176,0.5542483660130719}{55.1}{7.200000000000003}    & \cbox{0.33725490196078434,0.8549019607843137,0.40784313725490196}{62.1}{14.200000000000003}          & \cbox{0.33725490196078434,0.8549019607843137,0.40784313725490196}{\textbf{62.8}}{14.899999999999999} \\
    \midrule
    \multirow{2}{*}[-1mm]{\parbox[m][][c]{1.8cm}{\centering HHH}}  & LLaMA-7B  &             --                            & \cbox{0.9388081507112649,0.9408381391772395,0.9390849673202615}{59.3}{0.0} & \cbox{0.9492502883506344,0.897039600153787,0.897039600153787}{57.9}{-1.3999999999999986}   & \cbox{0.45093425605536336,0.8711418685121107,0.5082352941176471}{67.4}{8.100000000000009}  & \cbox{0.33725490196078434,0.8549019607843137,0.40784313725490196}{70.1}{10.799999999999997} & \cbox{0.33725490196078434,0.8549019607843137,0.40784313725490196}{\textbf{71.5}}{12.200000000000003} & \cbox{0.33725490196078434,0.8549019607843137,0.40784313725490196}{\textbf{71.5}}{12.200000000000003} \\[\sp]
                                                                   & LLaMA-13B &               --                          & \cbox{0.9388081507112649,0.9408381391772395,0.9390849673202615}{60.2}{0.0} & \cbox{0.9441753171856978,0.9247827758554401,0.9247827758554401}{59.7}{-0.5}                & \cbox{0.9547866205305652,0.8667743175701653,0.8667743175701653}{57.9}{-2.3000000000000043} & \cbox{0.3940945790080739,0.8630219146482122,0.4580392156862745}{\textbf{69.2}}{9.0}         & \cbox{0.8867051134179161,0.9333948481353326,0.8930718954248366}{61.1}{0.8999999999999986}            & \cbox{0.8867051134179161,0.9333948481353326,0.8930718954248366}{61.1}{0.8999999999999986}            \\
    \bottomrule
\end{tabular}

    \caption{Results of the different approaches across multiple tasks. We compute the accuracies on the test set of these benchmarks. \algo outperforms other approaches on most tasks. \algo performs well, even in cases where one of \generator or \discriminator is far worse than the other. \textbf{G}: Generative Ranking, \textbf{MI}: Mutual Information Ranking, \textbf{SC}: Self-Contrastive Ranking, \textbf{D}: Discriminative Ranking, \textbf{ER-G}: Equilibrium Ranking Generator, \textbf{ER-D}: Equilibrium Ranking Discriminator. * indicates the results from \cite{touvron2023llama}. Colors in the table entries are assigned relative to the G baseline, according to the colorbar \!\raisebox{-3.5mm}{\includegraphics[scale=.7]{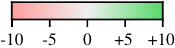}}\! (differences exceeding $\pm 10$ are clipped to $\pm 10$ when calculating the colors).\\[-4mm]}
    \label{tab:combined}
\end{table*}

\paragraph{MMLU}
The massive multi-task language understanding benchmark (MMLU) \citep{hendrycks2020measuring} is used to measure language model's multitask accuracy. It consists of questions in the multiple choice format across a wide variety of subdomains in social sciences, humanities and STEM. We evaluate our models in the zero-shot setting following the format described in \cite{hendrycks2020measuring, touvron2023llama} and the results are presented in the first row of \cref{tab:combined}. For both LLaMA-7B and LLaMA-13B, the \algo-based approaches matches or outperforms all other baselines. In fact, zero-shot LLaMA-7B with ER-D (39.9) outperforms 5-shot LLaMA-7B (35.1), while zero-shot LLaMA-13B with ER-D (45.1) is competitive with 5-shot LLaMA-13B (46.9). LLaMA-7B with ER-D (39.9) even outperforms zero-shot GPT3-175B (37.7) \citep{hendrycks2020measuring}, while zero-shot LLaMA-13B with ER-D (45.1) outperforms 5-shot GPT3-175B (43.9) \citep{hendrycks2020measuring}.

\paragraph{ARC}
The AI2 Reasoning Challenge (ARC) \citep{clark2018think} is an advanced question answering dataset used to study a model's knowledge and reasoning abilities based on grade school science questions. It is split in to two subcategories: easy (ARC-Easy) and challenge (ARC-Challenge). The challenge set was constructed as the set of questions that were answered incorrectly by retrieval and word co-occurence based algorithms. The results are presented in second and third rows of \cref{tab:combined}. On ARC-Easy, ER-D outperforms our implementation of generative ranking. We also note that \llamab with ER-D (76.4) outperform all the baseline approaches and is even competitive with the much larger PaLM-540B model (76.6) \citep{chowdhery2022palm}. On ARC-Challenge, ER-D significantly outperforms all the baseline approaches. We also note that \llamas with ER-D (58.3) and \llamab with ER-D (61.4) outperforms even the much larger models: LLaMA-65B (56.0) \citep{touvron2023llama} and PaLM-540B (53.0) \citep{chowdhery2022palm} by up to 8\%.

\paragraph{RACE}
RACE is a reading comprehension benchmark introduced in \cite{lai2017race} collected from English examinations of middle and high school students. The dataset is correspondingly split into RACE-middle and Race-high. The dataset consists of a passage followed by questions. The passages were constructed for evaluating student's English reasoning and understanding ability. The results on this benchmark is presented in rows 4 and 5 of \cref{tab:combined}. On RACE-middle, like before, ER-D based models outperforms all the baselines. We note that \llamab with ER-D (68.6) even outperforms much larger models: LLaMA-65B (67.9) \citep{touvron2023llama} and PaLM-540B (68.1) \citep{chowdhery2022palm}. On RACE-high, we have a similar story as with ARC-C. ER-D outperforms all baselines. \llamas with ER-D (56.4) is able to significantly outperform much larger models: LLaMA-65B (51.6) \citep{touvron2023llama} and PaLM-540B (49.1) \citep{chowdhery2022palm}.

\paragraph{HHH}
HHH (Helpful, Honest and Harmless) \citep{srivastava2023beyond, askell2021general} is a dataset of 200 multiple-choice designed to measure LM alignment with high-level quality guidelines. Here we use a different set of prompts for the \generator (see \cref{sec:hhh}). Results are presented in the last row of \cref{tab:combined}. \llamas with ER-D outperforms baselines; although \llamab with ER-D with the default parameter performs worse than discriminative ranking (D) (69.2), ER-D with $\lambda_\gen=0.01$ and $\lambda_\dis=1.0$ achieves an accuracy of 70.6\%.

\paragraph{TruthfulQA}
TruthfulQA \citep{lin2021truthfulqa} is a benchmark consisting of over 800 questions across a multitude of domains that were crafted to encourage humans to answer them incorrectly due to misconceptions. The dataset evaluates a model's ability to not generate false answers learnt from imitation learning on text. On this task, we consider \textbf{greedy decoding} in addition to our other ranking-based approaches. In this setting, 10 candidate sequences are sampled using nucleus and top-k sampling. These candidates are then ranked based on the approaches we described earlier. The results on the test set are presented in \cref{tab:truthfulqa}. Based on past work \citep{lin2021truthfulqa}, we measure BLEU accuracy (BLEU-Acc). For a sequence $a$, the BLEU-Acc over reference correct candidates $b_{\text{correct}}$ and reference incorrect candidates $b_{\text{correct}}$ is computed as follows:
\begin{align}
    \text{BLEU-Acc}(a) \coloneqq \mathbb{I}(\text{BLEU}(a, b_\text{correct}) > \text{BLEU}(a, b_\text{incorrect}))
\end{align}
Where $\text{BLEU}(a, b)$ computes the BLEU score \citep{papineni2002bleu} of a candidate string $a$ over a set of reference candidates $b$. With \llamas, we observe only modest improvements for ER-G and ER-D over the greedy baseline. However, with \llamab, we note increased scores for both methods. This benchmark is known to exhibit negative scaling \citep{lin2021truthfulqa} (performance drop as the model size increases). The performance difference with ER-G between \llamas and \llamab shows that \algo is in fact capable of mitigating this behavior.

\begin{table}[t]
    \def\cboxwgt{11mm}
    \def\rowheight{8.2mm}

    \centering
    \def\sp{2.6mm}
\begin{tabular}{llcccc|cc}
    \toprule
                                          &           &                                                                             &                                                                                                            &                                                                                                            &                                                                                                           & \multicolumn{2}{c}{\textbf{Equil. ranking}}                                                                                                                                                                                     \\
    Domain                                & Model     & Greedy                                                                      & MI                                                                                                         & SC                                                                                                         & D                                                                                                         & ER-G                                                                                                               & ER-D                                                                                                       \\
    \midrule
    \multirow{2}{*}[-2.5mm]{TruthfulQA}~~ & LLaMA-7B  & \cbox{0.9388081507112649,0.9408381391772395,0.9390849673202615}{33.41}{0.0} & \cbox{0.8582852748942714,0.9293348712033833,0.8679738562091504}{\pmstack{34.79}{0.90}}{1.3800000000000026} & \cbox{0.8488119953863899,0.9279815455594003,0.8596078431372549}{\pmstack{\textbf{34.91}}{0.57}}{1.5}       & \cbox{0.8961783929257978,0.9347481737793156,0.901437908496732}{\pmstack{34.17}{1.19}}{0.7600000000000051} & \cbox{0.867758554402153,0.9306881968473664,0.8763398692810458}{\pmstack{34.61}{0.99}}{1.2000000000000028}          & \cbox{0.8867051134179161,0.9333948481353326,0.8930718954248366}{\pmstack{34.27}{0.39}}{0.8600000000000065} \\[\sp]
                                          & LLaMA-13B & \cbox{0.9388081507112649,0.9408381391772395,0.9390849673202615}{33.05}{0.0} & \cbox{0.7446059207996925,0.9130949634755863,0.7675816993464052}{\pmstack{36.30}{0.37}}{3.25}               & \cbox{0.8488119953863899,0.9279815455594003,0.8596078431372549}{\pmstack{34.61}{1.33}}{1.5600000000000023} & \cbox{0.5788235294117647,0.8894117647058823,0.6211764705882352}{\pmstack{39.05}{1.42}}{6.0}               & \cbox{0.5314571318723568,0.8826451364859669,0.5793464052287581}{\pmstack{\textbf{39.83}}{2.20}}{6.780000000000001} & \cbox{0.6025067281814687,0.8927950788158401,0.6420915032679738}{\pmstack{38.63}{1.76}}{5.580000000000005}  \\
    \bottomrule
\end{tabular}

    \caption{Results on TruthfulQA (Generative). Average BLEU-Acc results on the held-out set across 5 runs. \llamab with ER-G outperforms or is on par with all baselines. \textbf{MI}: Mutual Information Ranking, \textbf{SC}: Self-Contrastive Ranking, \textbf{D}: Discriminative Ranking, \textbf{ER-G}: Equilibrium Ranking Generator, \textbf{ER-D}: Equilibrium Ranking Discriminator. $\pm$ indicates 1 standard deviation computed across 5 runs. Colors are as in \cref{tab:combined}, relative to the Greedy baseline.}
    \label{tab:truthfulqa}
\end{table}

\paragraph{GSM8K}
\label{sec:gsm}
In our last set of experiments, we consider grade-school math (GSM8K) \citep{cobbe2021training}, a popular benchmark used to study model's mathematical reasoning ability. We use this benchmark to study whether we can combine our approach with chain-of-thought \citep{wei2022chain}. As we described earlier, we generate 20 candidate reasoning paths sampled using nucleus and top-k using the 8-shot setup proposed in \citet{wei2022chain}. We employ self-consistency \citep{wang2023selfconsistency} over the candidate sequences, where we score each reasoning path with our baselines. Finally, we aggregate the scores for each answer corresponding to the reasoning paths and pick the answer with the highest score. In \cref{tab:gsm8k}, we present the results. We note that \algo-based approaches performs on par or slightly better compared to self-consistency (majority vote).

\begin{table}[t]
    \def\cboxwgt{9.6mm}
    \def\rowheight{8.2mm}
    
    \centering
    \def\sp{2.6mm}
\begin{tabular}{llccccc|cc}
    \toprule
                                     &           &                                                                            &                                                                                                         &                                                                                                        &                                                                                                                  &                                                                                                         & \multicolumn{2}{c}{\textbf{Equil. ranking}}                                                                                                                                                                                 \\
    Domain                           & Model     & Greedy                                                                     & MV                                                                                                      & MI                                                                                                     & SC                                                                                                               & D                                                                                                       & ER-G                                                                                                     & ER-D                                                                                                             \\
    \midrule
    \multirow{2}{*}[-2.5mm]{GSM8K}~~ & LLaMA-7B  & \cbox{0.9388081507112649,0.9408381391772395,0.9390849673202615}{10.8}{0.0} & \cbox{0.7067128027681661,0.907681660899654,0.7341176470588235}{\pmstack{14.7}{0.2}}{3.8999999999999986} & \cbox{0.711449442522107,0.9083583237216455,0.7383006535947713}{\pmstack{14.6}{0.5}}{3.799999999999999} & \cbox{0.7824990388312187,0.9185082660515186,0.801045751633987}{\pmstack{13.4}{0.3}}{2.5999999999999996}          & \cbox{0.6877662437524029,0.9049750096116878,0.7173856209150327}{\pmstack{15.0}{0.6}}{4.199999999999999} & \cbox{0.8061822376009227,0.9218915801614763,0.8219607843137254}{\pmstack{13.0}{0.5}}{2.1999999999999993} & \cbox{0.6782929642445213,0.9036216839677047,0.7090196078431372}{\pmstack{\textbf{15.1}}{0.6}}{4.299999999999999} \\[\sp]
                                     & LLaMA-13B & \cbox{0.9388081507112649,0.9408381391772395,0.9390849673202615}{14.9}{0.0} & \cbox{0.4793540945790081,0.87520184544406,0.5333333333333334}{\pmstack{22.5}{0.5}}{7.6}                 & \cbox{0.4793540945790081,0.87520184544406,0.5333333333333334}{\pmstack{22.5}{0.8}}{7.6}                & \cbox{0.4461976163014226,0.8704652056901192,0.5040522875816993}{\pmstack{\textbf{23.1}}{0.5}}{8.200000000000001} & \cbox{0.4793540945790081,0.87520184544406,0.5333333333333334}{\pmstack{22.5}{0.6}}{7.6}                 & \cbox{0.4793540945790081,0.87520184544406,0.5333333333333334}{\pmstack{22.5}{0.6}}{7.6}                  & \cbox{0.45093425605536336,0.8711418685121107,0.5082352941176471}{\pmstack{23.0}{0.5}}{8.1}                       \\
    \bottomrule
\end{tabular}

    \caption{Average accuracy of methods on the test set of GSM8K acroos 5 runs. In all cases, except greedy, 20 candidates were sampled.  \algo-based approaches performs on par or slightly better compared to the majority vote baseline. \textbf{MV}: Majority Vote, \textbf{MI}: Mutual Information Ranking, \textbf{SC}: Self-Contrastive Ranking, \textbf{D}: Discriminative Ranking, \textbf{ER-G}: Equilibrium Ranking Generator, \textbf{ER-D}: Equilibrium Ranking Discriminator. $\pm$ indicates 1 standard deviation. Colors are as in \cref{tab:combined}, relative to the Greedy basline.}
    \label{tab:gsm8k}
\end{table}

\paragraph{Discussion}
The application of \algo-based approaches consistently yields improved results, surpassing or at least matching the performance of all baseline approaches across various tasks. This robustness is particularly interesting, as it demonstrates that \algo is adept at handling diverse scenarios, even in situations when the initial \generator or \discriminator are not effective. As \algo is a sampling strategy, it can even be combined with deliberation methods like self-consistency \citep{wang2023selfconsistency} or tree-of-thought \citep{yao2023tree}. Finally, we note that \algo demonstrates computational efficiency by eliminating the need for repetitive queries to language models.
\section{Other Related Work}
Many decoding strategies have been proposed for language models, such as top-k sampling~\citep{fan-etal-2018-hierarchical}, nucleus sampling~\citep{Holtzman2020The}, and typical sampling~\citep{meister2023locally}. These methods primarily focus on producing diverse, high-quality text from a language model. However, they decode from the LM without any emphasis on the correctness of the generated sequences. As we show in \cref{sec:results}, \algo is naturally complementary and be combined with any of these sampling strategies.

Re-ranking is a common approach for selecting the correct answer from a set of candidates sampled from LM. \cite{cobbe2021training} train a verifier to re-ranked the sampled outputs. \cite{shen-etal-2021-generate-rank} jointly train a ranking model with the generation model to improve the model accuracy. \cite{thoppilan2022lamda} collect additional human annotations to train the ranking model for response filtering. As we discuss in \cref{sec:formulation}, our work focuses on leveraging an existing LM and using them in a training-free manner as a discriminator. However, we note that we do not make any specific assumption on the specific form of a the \generator or \discriminator. As such, \algo can be combined with these approaches.

As previously mentioned, \algo differs from recent deliberation methods, as highlighted in various recent work \citep{wei2022chain, madaan2023selfrefine, shinn2023reflexion, yao2023tree, dohan2022language}. In \cref{sec:results}, we demonstrate how \algo can be integrated with these approaches. In another line of work, \cite{du2023improving} and \citet{chen2023reconcile} employ multiple instances of language models suggest and "debate" individual responses and reasoning processes across multiple iterations, ultimately converging on a shared final answer. In contrast, \algo can be viewed as a variant of this multi-agent debate, wherein the "debate" occurs within the regret-minimization framework rather than in the context of language models.

\section{Conclusion}
We have presented \algo, a training-free, game theoretic approach for generating from language models (LMs).
\algo reconciles scores from generative and discriminative LM decoding procedures by formulating decoding as an imperfect-information signaling game between a \generator and a \discriminator, %
and leveraging computational game solving techniques to compute approximate equilibria of this game. When applied to 6 diverse question answering benchmarks: MMLU, ARC, RACE, HHH, TruthfulQA and, GSM8K, \algo offers substantial improvements over existing generative, discriminative, and mixed decoding procedures: applying \algo to \llamas sometimes outperforms much larger LLaMA-65B and PaLM-540B models. These results highlight the usefulness of game-theoretic tools in formalizing desiderata like truthfulness and stability in language modeling. Beyond the applications studied here (which focus mainly on question answer), future work might apply this toolkit to more general tasks like long-form text generation.


\begin{thebibliography}{50}
\providecommand{\natexlab}[1]{#1}
\providecommand{\url}[1]{\texttt{#1}}
\expandafter\ifx\csname urlstyle\endcsname\relax
  \providecommand{\doi}[1]{doi: #1}\else
  \providecommand{\doi}{doi: \begingroup \urlstyle{rm}\Url}\fi

\bibitem[Anagnostides et~al.(2022)Anagnostides, Panageas, Farina, and Sandholm]{Anagnostides22:Last-Iterate}
Ioannis Anagnostides, Ioannis Panageas, Gabriele Farina, and Tuomas Sandholm.
\newblock On last-iterate convergence beyond zero-sum games.
\newblock In \emph{International Conference on Machine Learning}, 2022.

\bibitem[Askell et~al.(2021)Askell, Bai, Chen, Drain, Ganguli, Henighan, Jones, Joseph, Mann, DasSarma, et~al.]{askell2021general}
Amanda Askell, Yuntao Bai, Anna Chen, Dawn Drain, Deep Ganguli, Tom Henighan, Andy Jones, Nicholas Joseph, Ben Mann, Nova DasSarma, et~al.
\newblock A general language assistant as a laboratory for alignment.
\newblock \emph{arXiv preprint arXiv:2112.00861}, 2021.

\bibitem[Bakhtin et~al.(2023)Bakhtin, Wu, Lerer, Gray, Jacob, Farina, Miller, and Brown]{bakhtin2022mastering}
Anton Bakhtin, David~J Wu, Adam Lerer, Jonathan Gray, Athul~Paul Jacob, Gabriele Farina, Alexander~H Miller, and Noam Brown.
\newblock Mastering the game of no-press {D}iplomacy via human-regularized reinforcement learning and planning.
\newblock In \emph{The Eleventh International Conference on Learning Representations}, 2023.

\bibitem[Brown \& Sandholm(2018)Brown and Sandholm]{brown2018superhuman}
Noam Brown and Tuomas Sandholm.
\newblock Superhuman ai for heads-up no-limit poker: Libratus beats top professionals.
\newblock \emph{Science}, 359\penalty0 (6374):\penalty0 418--424, 2018.

\bibitem[Brown \& Sandholm(2019)Brown and Sandholm]{brown2019superhuman}
Noam Brown and Tuomas Sandholm.
\newblock Superhuman ai for multiplayer poker.
\newblock \emph{Science}, 365\penalty0 (6456):\penalty0 885--890, 2019.

\bibitem[Brown et~al.(2020)Brown, Mann, Ryder, Subbiah, Kaplan, Dhariwal, Neelakantan, Shyam, Sastry, Askell, et~al.]{brown2020language}
Tom Brown, Benjamin Mann, Nick Ryder, Melanie Subbiah, Jared~D Kaplan, Prafulla Dhariwal, Arvind Neelakantan, Pranav Shyam, Girish Sastry, Amanda Askell, et~al.
\newblock Language models are few-shot learners.
\newblock \emph{Advances in neural information processing systems}, 33:\penalty0 1877--1901, 2020.

\bibitem[Chen et~al.(2023{\natexlab{a}})Chen, Saha, and Bansal]{chen2023reconcile}
Justin Chih-Yao Chen, Swarnadeep Saha, and Mohit Bansal.
\newblock Reconcile: Round-table conference improves reasoning via consensus among diverse llms.
\newblock \emph{arXiv preprint arXiv:2309.13007}, 2023{\natexlab{a}}.

\bibitem[Chen et~al.(2023{\natexlab{b}})Chen, Lin, Sch{\"a}rli, and Zhou]{chen2023teaching}
Xinyun Chen, Maxwell Lin, Nathanael Sch{\"a}rli, and Denny Zhou.
\newblock Teaching large language models to self-debug.
\newblock \emph{arXiv preprint arXiv:2304.05128}, 2023{\natexlab{b}}.

\bibitem[Chen et~al.(2022)Chen, Yuan, Cui, Liu, and Ji]{chen2022close}
Yangyi Chen, Lifan Yuan, Ganqu Cui, Zhiyuan Liu, and Heng Ji.
\newblock A close look into the calibration of pre-trained language models.
\newblock \emph{arXiv preprint arXiv:2211.00151}, 2022.

\bibitem[Chowdhery et~al.(2022)Chowdhery, Narang, Devlin, Bosma, Mishra, Roberts, Barham, Chung, Sutton, Gehrmann, et~al.]{chowdhery2022palm}
Aakanksha Chowdhery, Sharan Narang, Jacob Devlin, Maarten Bosma, Gaurav Mishra, Adam Roberts, Paul Barham, Hyung~Won Chung, Charles Sutton, Sebastian Gehrmann, et~al.
\newblock Palm: Scaling language modeling with pathways.
\newblock \emph{arXiv preprint arXiv:2204.02311}, 2022.

\bibitem[Clark et~al.(2018)Clark, Cowhey, Etzioni, Khot, Sabharwal, Schoenick, and Tafjord]{clark2018think}
Peter Clark, Isaac Cowhey, Oren Etzioni, Tushar Khot, Ashish Sabharwal, Carissa Schoenick, and Oyvind Tafjord.
\newblock Think you have solved question answering? try arc, the ai2 reasoning challenge.
\newblock \emph{arXiv preprint arXiv:1803.05457}, 2018.

\bibitem[Cobbe et~al.(2021)Cobbe, Kosaraju, Bavarian, Chen, Jun, Kaiser, Plappert, Tworek, Hilton, Nakano, Hesse, and Schulman]{cobbe2021training}
Karl Cobbe, Vineet Kosaraju, Mohammad Bavarian, Mark Chen, Heewoo Jun, Lukasz Kaiser, Matthias Plappert, Jerry Tworek, Jacob Hilton, Reiichiro Nakano, Christopher Hesse, and John Schulman.
\newblock Training verifiers to solve math word problems, 2021.

\bibitem[Dohan et~al.(2022)Dohan, Xu, Lewkowycz, Austin, Bieber, Lopes, Wu, Michalewski, Saurous, Sohl-Dickstein, et~al.]{dohan2022language}
David Dohan, Winnie Xu, Aitor Lewkowycz, Jacob Austin, David Bieber, Raphael~Gontijo Lopes, Yuhuai Wu, Henryk Michalewski, Rif~A Saurous, Jascha Sohl-Dickstein, et~al.
\newblock Language model cascades.
\newblock \emph{arXiv preprint arXiv:2207.10342}, 2022.

\bibitem[Du et~al.(2023)Du, Li, Torralba, Tenenbaum, and Mordatch]{du2023improving}
Yilun Du, Shuang Li, Antonio Torralba, Joshua~B Tenenbaum, and Igor Mordatch.
\newblock Improving factuality and reasoning in language models through multiagent debate.
\newblock \emph{arXiv preprint arXiv:2305.14325}, 2023.

\bibitem[FAIR et~al.(2022)FAIR, Bakhtin, Brown, Dinan, Farina, Flaherty, Fried, Goff, Gray, Hu, et~al.]{meta2022human}
Meta FAIR, Anton Bakhtin, Noam Brown, Emily Dinan, Gabriele Farina, Colin Flaherty, Daniel Fried, Andrew Goff, Jonathan Gray, Hengyuan Hu, et~al.
\newblock Human-level play in the game of diplomacy by combining language models with strategic reasoning.
\newblock \emph{Science}, 378\penalty0 (6624):\penalty0 1067--1074, 2022.

\bibitem[Fan et~al.(2018{\natexlab{a}})Fan, Lewis, and Dauphin]{fan-etal-2018-hierarchical}
Angela Fan, Mike Lewis, and Yann Dauphin.
\newblock Hierarchical neural story generation.
\newblock In \emph{Proceedings of the 56th Annual Meeting of the Association for Computational Linguistics (Volume 1: Long Papers)}, pp.\  889--898, Melbourne, Australia, July 2018{\natexlab{a}}. Association for Computational Linguistics.
\newblock \doi{10.18653/v1/P18-1082}.
\newblock URL \url{https://aclanthology.org/P18-1082}.

\bibitem[Fan et~al.(2018{\natexlab{b}})Fan, Lewis, and Dauphin]{topk}
Angela Fan, Mike Lewis, and Yann Dauphin.
\newblock Hierarchical neural story generation.
\newblock In \emph{Proceedings of the 56th Annual Meeting of the Association for Computational Linguistics (Volume 1: Long Papers)}, pp.\  889--898, 2018{\natexlab{b}}.

\bibitem[Farina et~al.(2019)Farina, Kroer, and Sandholm]{Farina19:Online}
Gabriele Farina, Christian Kroer, and Tuomas Sandholm.
\newblock Online convex optimization for sequential decision processes and extensive-form games.
\newblock In \emph{AAAI Conference on Artificial Intelligence (AAAI)}, 2019.

\bibitem[Franke(2013)]{franke2013game}
Michael Franke.
\newblock Game theoretic pragmatics.
\newblock \emph{Philosophy Compass}, 8\penalty0 (3):\penalty0 269--284, 2013.

\bibitem[Franke(2017)]{franke2017game}
Michael Franke.
\newblock Game theory in pragmatics: Evolution, rationality, and reasoning.
\newblock In \emph{Oxford Research Encyclopedia of Linguistics}. 2017.

\bibitem[Ganguli et~al.(2023)Ganguli, Askell, Schiefer, Liao, Lukošiūtė, Chen, Goldie, Mirhoseini, Olsson, Hernandez, Drain, Li, Tran-Johnson, Perez, Kernion, Kerr, Mueller, Landau, Ndousse, Nguyen, Lovitt, Sellitto, Elhage, Mercado, DasSarma, Rausch, Lasenby, Larson, Ringer, Kundu, Kadavath, Johnston, Kravec, Showk, Lanham, Telleen-Lawton, Henighan, Hume, Bai, Hatfield-Dodds, Mann, Amodei, Joseph, McCandlish, Brown, Olah, Clark, Bowman, and Kaplan]{ganguli2023capacity}
Deep Ganguli, Amanda Askell, Nicholas Schiefer, Thomas~I. Liao, Kamilė Lukošiūtė, Anna Chen, Anna Goldie, Azalia Mirhoseini, Catherine Olsson, Danny Hernandez, Dawn Drain, Dustin Li, Eli Tran-Johnson, Ethan Perez, Jackson Kernion, Jamie Kerr, Jared Mueller, Joshua Landau, Kamal Ndousse, Karina Nguyen, Liane Lovitt, Michael Sellitto, Nelson Elhage, Noemi Mercado, Nova DasSarma, Oliver Rausch, Robert Lasenby, Robin Larson, Sam Ringer, Sandipan Kundu, Saurav Kadavath, Scott Johnston, Shauna Kravec, Sheer~El Showk, Tamera Lanham, Timothy Telleen-Lawton, Tom Henighan, Tristan Hume, Yuntao Bai, Zac Hatfield-Dodds, Ben Mann, Dario Amodei, Nicholas Joseph, Sam McCandlish, Tom Brown, Christopher Olah, Jack Clark, Samuel~R. Bowman, and Jared Kaplan.
\newblock The capacity for moral self-correction in large language models, 2023.

\bibitem[Glaese et~al.(2022)Glaese, McAleese, Trebacz, Aslanides, Firoiu, Ewalds, Rauh, Weidinger, Chadwick, Thacker, et~al.]{glaese2022improving}
Amelia Glaese, Nat McAleese, Maja Trebacz, John Aslanides, Vlad Firoiu, Timo Ewalds, Maribeth Rauh, Laura Weidinger, Martin Chadwick, Phoebe Thacker, et~al.
\newblock Improving alignment of dialogue agents via targeted human judgements.
\newblock \emph{arXiv preprint arXiv:2209.14375}, 2022.

\bibitem[Han et~al.(2022)Han, Hao, Dong, Sun, and Wei]{han2022prototypical}
Zhixiong Han, Yaru Hao, Li~Dong, Yutao Sun, and Furu Wei.
\newblock Prototypical calibration for few-shot learning of language models.
\newblock In \emph{The Eleventh International Conference on Learning Representations}, 2022.

\bibitem[Hendrycks et~al.(2020)Hendrycks, Burns, Basart, Zou, Mazeika, Song, and Steinhardt]{hendrycks2020measuring}
Dan Hendrycks, Collin Burns, Steven Basart, Andy Zou, Mantas Mazeika, Dawn Song, and Jacob Steinhardt.
\newblock Measuring massive multitask language understanding.
\newblock In \emph{International Conference on Learning Representations}, 2020.

\bibitem[Holtzman et~al.(2019)Holtzman, Buys, Du, Forbes, and Choi]{nucleus}
Ari Holtzman, Jan Buys, Li~Du, Maxwell Forbes, and Yejin Choi.
\newblock The curious case of neural text degeneration.
\newblock In \emph{International Conference on Learning Representations}, 2019.

\bibitem[Holtzman et~al.(2020)Holtzman, Buys, Du, Forbes, and Choi]{Holtzman2020The}
Ari Holtzman, Jan Buys, Li~Du, Maxwell Forbes, and Yejin Choi.
\newblock The curious case of neural text degeneration.
\newblock In \emph{International Conference on Learning Representations}, 2020.
\newblock URL \url{https://openreview.net/forum?id=rygGQyrFvH}.

\bibitem[Jacob et~al.(2022)Jacob, Wu, Farina, Lerer, Hu, Bakhtin, Andreas, and Brown]{Jacob22:Modeling}
Athul~Paul Jacob, David~J. Wu, Gabriele Farina, Adam Lerer, Hengyuan Hu, Anton Bakhtin, Jacob Andreas, and Noam Brown.
\newblock Modeling strong and human-like gameplay with kl-regularized search.
\newblock In \emph{International Conference on Machine Learning}, 2022.

\bibitem[Jiang et~al.(2020)Jiang, Xu, Araki, and Neubig]{jiang2020can}
Zhengbao Jiang, Frank~F Xu, Jun Araki, and Graham Neubig.
\newblock How can we know what language models know?
\newblock \emph{Transactions of the Association for Computational Linguistics}, 8:\penalty0 423--438, 2020.

\bibitem[Lai et~al.(2017)Lai, Xie, Liu, Yang, and Hovy]{lai2017race}
Guokun Lai, Qizhe Xie, Hanxiao Liu, Yiming Yang, and Eduard Hovy.
\newblock Race: Large-scale reading comprehension dataset from examinations.
\newblock In \emph{Proceedings of the 2017 Conference on Empirical Methods in Natural Language Processing}, pp.\  785--794, 2017.

\bibitem[Lewis(2008)]{lewis2008convention}
David Lewis.
\newblock \emph{Convention: A philosophical study}.
\newblock John Wiley \& Sons, 2008.

\bibitem[Li \& Jurafsky(2016)Li and Jurafsky]{li2016mutual}
Jiwei Li and Dan Jurafsky.
\newblock Mutual information and diverse decoding improve neural machine translation.
\newblock \emph{arXiv preprint arXiv:1601.00372}, 2016.

\bibitem[Li et~al.(2022)Li, Holtzman, Fried, Liang, Eisner, Hashimoto, Zettlemoyer, and Lewis]{li2022contrastive}
Xiang~Lisa Li, Ari Holtzman, Daniel Fried, Percy Liang, Jason Eisner, Tatsunori Hashimoto, Luke Zettlemoyer, and Mike Lewis.
\newblock Contrastive decoding: Open-ended text generation as optimization.
\newblock \emph{arXiv preprint arXiv:2210.15097}, 2022.

\bibitem[Lin et~al.(2022)Lin, Hilton, and Evans]{lin2021truthfulqa}
Stephanie Lin, Jacob Hilton, and Owain Evans.
\newblock Truthfulqa: Measuring how models mimic human falsehoods.
\newblock In \emph{Proceedings of the 60th Annual Meeting of the Association for Computational Linguistics (Volume 1: Long Papers)}, pp.\  3214--3252, 2022.

\bibitem[Madaan et~al.(2023)Madaan, Tandon, Gupta, Hallinan, Gao, Wiegreffe, Alon, Dziri, Prabhumoye, Yang, Gupta, Majumder, Hermann, Welleck, Yazdanbakhsh, and Clark]{madaan2023selfrefine}
Aman Madaan, Niket Tandon, Prakhar Gupta, Skyler Hallinan, Luyu Gao, Sarah Wiegreffe, Uri Alon, Nouha Dziri, Shrimai Prabhumoye, Yiming Yang, Shashank Gupta, Bodhisattwa~Prasad Majumder, Katherine Hermann, Sean Welleck, Amir Yazdanbakhsh, and Peter Clark.
\newblock Self-refine: Iterative refinement with self-feedback, 2023.

\bibitem[McKenzie et~al.(2023)McKenzie, Lyzhov, Pieler, Parrish, Mueller, Prabhu, McLean, Kirtland, Ross, Liu, et~al.]{mckenzie2023inverse}
Ian~R McKenzie, Alexander Lyzhov, Michael Pieler, Alicia Parrish, Aaron Mueller, Ameya Prabhu, Euan McLean, Aaron Kirtland, Alexis Ross, Alisa Liu, et~al.
\newblock Inverse scaling: When bigger isn't better.
\newblock \emph{arXiv preprint arXiv:2306.09479}, 2023.

\bibitem[Meister et~al.(2023)Meister, Pimentel, Wiher, and Cotterell]{meister2023locally}
Clara~Isabel Meister, Tiago Pimentel, Gian Wiher, and Ryan Cotterell.
\newblock Locally typical sampling.
\newblock \emph{Transactions of the Association for Computational Linguistics}, 11:\penalty0 102--121, 2023.

\bibitem[Mitchell et~al.(2022)Mitchell, Noh, Li, Armstrong, Agarwal, Liu, Finn, and Manning]{mitchell2022enhancing}
Eric Mitchell, Joseph Noh, Siyan Li, Will Armstrong, Ananth Agarwal, Patrick Liu, Chelsea Finn, and Christopher~D Manning.
\newblock Enhancing self-consistency and performance of pre-trained language models through natural language inference.
\newblock In \emph{Proceedings of the 2022 Conference on Empirical Methods in Natural Language Processing}, pp.\  1754--1768, 2022.

\bibitem[Monderer \& Shapley(1996)Monderer and Shapley]{Monderer96:Potential}
Dov Monderer and Lloyd~S. Shapley.
\newblock Potential games.
\newblock \emph{Games and Economic Behavior}, 1\penalty0 (14):\penalty0 124--143, 1996.

\bibitem[Ouyang et~al.(2022)Ouyang, Wu, Jiang, Almeida, Wainwright, Mishkin, Zhang, Agarwal, Slama, Ray, et~al.]{ouyang2022training}
Long Ouyang, Jeffrey Wu, Xu~Jiang, Diogo Almeida, Carroll Wainwright, Pamela Mishkin, Chong Zhang, Sandhini Agarwal, Katarina Slama, Alex Ray, et~al.
\newblock Training language models to follow instructions with human feedback.
\newblock \emph{Advances in Neural Information Processing Systems}, 35:\penalty0 27730--27744, 2022.

\bibitem[Papineni et~al.(2002)Papineni, Roukos, Ward, and Zhu]{papineni2002bleu}
Kishore Papineni, Salim Roukos, Todd Ward, and Wei-Jing Zhu.
\newblock Bleu: a method for automatic evaluation of machine translation.
\newblock In \emph{Proceedings of the 40th annual meeting of the Association for Computational Linguistics}, pp.\  311--318, 2002.

\bibitem[Perolat et~al.(2022)Perolat, De~Vylder, Hennes, Tarassov, Strub, de~Boer, Muller, Connor, Burch, Anthony, et~al.]{perolat2022mastering}
Julien Perolat, Bart De~Vylder, Daniel Hennes, Eugene Tarassov, Florian Strub, Vincent de~Boer, Paul Muller, Jerome~T Connor, Neil Burch, Thomas Anthony, et~al.
\newblock Mastering the game of stratego with model-free multiagent reinforcement learning.
\newblock \emph{Science}, 378\penalty0 (6623):\penalty0 990--996, 2022.

\bibitem[Shen et~al.(2021)Shen, Yin, Li, Shang, Jiang, Zhang, and Liu]{shen-etal-2021-generate-rank}
Jianhao Shen, Yichun Yin, Lin Li, Lifeng Shang, Xin Jiang, Ming Zhang, and Qun Liu.
\newblock Generate {\&} rank: A multi-task framework for math word problems.
\newblock In \emph{Findings of the Association for Computational Linguistics: EMNLP 2021}, pp.\  2269--2279, Punta Cana, Dominican Republic, November 2021. Association for Computational Linguistics.
\newblock \doi{10.18653/v1/2021.findings-emnlp.195}.
\newblock URL \url{https://aclanthology.org/2021.findings-emnlp.195}.

\bibitem[Shinn et~al.(2023)Shinn, Cassano, Labash, Gopinath, Narasimhan, and Yao]{shinn2023reflexion}
Noah Shinn, Federico Cassano, Beck Labash, Ashwin Gopinath, Karthik Narasimhan, and Shunyu Yao.
\newblock Reflexion: Language agents with verbal reinforcement learning, 2023.

\bibitem[Srivastava et~al.(2023)Srivastava, Rastogi, Rao, Shoeb, Abid, Fisch, Brown, Santoro, Gupta, Garriga-Alonso, et~al.]{srivastava2023beyond}
Aarohi Srivastava, Abhinav Rastogi, Abhishek Rao, Abu Awal~Md Shoeb, Abubakar Abid, Adam Fisch, Adam~R Brown, Adam Santoro, Aditya Gupta, Adri{\`a} Garriga-Alonso, et~al.
\newblock Beyond the imitation game: Quantifying and extrapolating the capabilities of language models.
\newblock \emph{Transactions on Machine Learning Research}, 2023.

\bibitem[Thoppilan et~al.(2022)Thoppilan, Freitas, Hall, Shazeer, Kulshreshtha, Cheng, Jin, Bos, Baker, Du, Li, Lee, Zheng, Ghafouri, Menegali, Huang, Krikun, Lepikhin, Qin, Chen, Xu, Chen, Roberts, Bosma, Zhao, Zhou, Chang, Krivokon, Rusch, Pickett, Srinivasan, Man, Meier-Hellstern, Morris, Doshi, Santos, Duke, Soraker, Zevenbergen, Prabhakaran, Diaz, Hutchinson, Olson, Molina, Hoffman-John, Lee, Aroyo, Rajakumar, Butryna, Lamm, Kuzmina, Fenton, Cohen, Bernstein, Kurzweil, Aguera-Arcas, Cui, Croak, Chi, and Le]{thoppilan2022lamda}
Romal Thoppilan, Daniel~De Freitas, Jamie Hall, Noam Shazeer, Apoorv Kulshreshtha, Heng-Tze Cheng, Alicia Jin, Taylor Bos, Leslie Baker, Yu~Du, YaGuang Li, Hongrae Lee, Huaixiu~Steven Zheng, Amin Ghafouri, Marcelo Menegali, Yanping Huang, Maxim Krikun, Dmitry Lepikhin, James Qin, Dehao Chen, Yuanzhong Xu, Zhifeng Chen, Adam Roberts, Maarten Bosma, Vincent Zhao, Yanqi Zhou, Chung-Ching Chang, Igor Krivokon, Will Rusch, Marc Pickett, Pranesh Srinivasan, Laichee Man, Kathleen Meier-Hellstern, Meredith~Ringel Morris, Tulsee Doshi, Renelito~Delos Santos, Toju Duke, Johnny Soraker, Ben Zevenbergen, Vinodkumar Prabhakaran, Mark Diaz, Ben Hutchinson, Kristen Olson, Alejandra Molina, Erin Hoffman-John, Josh Lee, Lora Aroyo, Ravi Rajakumar, Alena Butryna, Matthew Lamm, Viktoriya Kuzmina, Joe Fenton, Aaron Cohen, Rachel Bernstein, Ray Kurzweil, Blaise Aguera-Arcas, Claire Cui, Marian Croak, Ed~Chi, and Quoc Le.
\newblock Lamda: Language models for dialog applications, 2022.

\bibitem[Touvron et~al.(2023)Touvron, Lavril, Izacard, Martinet, Lachaux, Lacroix, Rozi{\`e}re, Goyal, Hambro, Azhar, et~al.]{touvron2023llama}
Hugo Touvron, Thibaut Lavril, Gautier Izacard, Xavier Martinet, Marie-Anne Lachaux, Timoth{\'e}e Lacroix, Baptiste Rozi{\`e}re, Naman Goyal, Eric Hambro, Faisal Azhar, et~al.
\newblock Llama: Open and efficient foundation language models.
\newblock \emph{arXiv preprint arXiv:2302.13971}, 2023.

\bibitem[Wang et~al.(2022)Wang, Wei, Schuurmans, Le, Chi, Narang, Chowdhery, and Zhou]{wang2023selfconsistency}
Xuezhi Wang, Jason Wei, Dale Schuurmans, Quoc~V Le, Ed~H Chi, Sharan Narang, Aakanksha Chowdhery, and Denny Zhou.
\newblock Self-consistency improves chain of thought reasoning in language models.
\newblock In \emph{The Eleventh International Conference on Learning Representations}, 2022.

\bibitem[Wei et~al.(2022)Wei, Wang, Schuurmans, Bosma, Xia, Chi, Le, Zhou, et~al.]{wei2022chain}
Jason Wei, Xuezhi Wang, Dale Schuurmans, Maarten Bosma, Fei Xia, Ed~Chi, Quoc~V Le, Denny Zhou, et~al.
\newblock Chain-of-thought prompting elicits reasoning in large language models.
\newblock \emph{Advances in Neural Information Processing Systems}, 35:\penalty0 24824--24837, 2022.

\bibitem[Yao et~al.(2023)Yao, Yu, Zhao, Shafran, Griffiths, Cao, and Narasimhan]{yao2023tree}
Shunyu Yao, Dian Yu, Jeffrey Zhao, Izhak Shafran, Thomas~L Griffiths, Yuan Cao, and Karthik Narasimhan.
\newblock Tree of thoughts: Deliberate problem solving with large language models.
\newblock \emph{arXiv preprint arXiv:2305.10601}, 2023.

\bibitem[Zinkevich et~al.(2007)Zinkevich, Bowling, Johanson, and Piccione]{Zinkevich07:Regret}
Martin Zinkevich, Michael Bowling, Michael Johanson, and Carmelo Piccione.
\newblock Regret minimization in games with incomplete information.
\newblock In \emph{Neural Information Processing Systems (NIPS)}, 2007.

\end{thebibliography}


\newpage
\appendix
\section{Details about Regret and Regret Decomposition Methods}
\label{app:cfr}

After each repetition $t$ of the game, each player---in this case, the \generator and the \discriminator---refines their policies, in such a way that throughout the course of time, the \emph{regrets}
\begin{align}
    \mathrm{Reg}^{(T)}_G & \defeq \max_{\pi^*_G} \left\{ \sum_{t=1}^T u_G(\pi^*_G, \pi^{(t)}_D) -  \sum_{t=1}^T u_G(\pi^{(t)}_G, \pi^{(t)}_D)\right\}, \label{eq:reg G} \\
    \mathrm{Reg}^{(T)}_D & \defeq \max_{\pi^*_D} \left\{ \sum_{t=1}^T u_D(\pi^{(t)}_G, \pi^*_D) -  \sum_{t=1}^T u_D(\pi^{(t)}_G, \pi^{(t)}_D)\right\},
\end{align}
cumulated by the players are guaranteed to grow sublinearly as a function of the number of rounds of learning $T$.

As mentioned in the body, a classic observation in the theory of imperfect-information sequential games is that minimization of regret (viewed as a function of their overall policy on the game tree) can be achieved by solving separate, \emph{local}, regret minimization problems at each information set (\textit{i.e.}, decision point) of the game. In our case, these techniques enable us to decompose the policy update of the players into separate updates for each \truthobj $v$ (for the \generator) and for each sequence $y$ (for the \discriminator).
More specifically, suppose that the \generator updates their policies $\pi_G^{(t)}(\cdot \mid \query, v)$ independently for each \truthobj $v\in\{\correct,\incorrect\}$ they might receive, seeking to independently minimize regret
\[
    \mathrm{Reg}^{(T)}_G(v) \defeq \max_{\pi^* \in \Delta(\sequences)} \left\{\sum_{t=1}^T \tilde u^{(t)}_G(\pi^* \mid \query,v) - \tilde u^{(t)}_G(\pi_G^{(t)}(\cdot\mid \query,v) \mid \query, v)\right\}
\]
with respect to the following \emph{counterfactual utility functions}
\begin{align}
    \tilde u^{(t)}_G(\pi_G \mid \query, v) \defeq -\lambda_G \mathrm{D}_\mathrm{KL}\bigg(\pi_G(\cdot \mid \query, v) \,\bigg\|\, \pi_G^{(0)}(\cdot \mid \query, v) \bigg) + \frac{1}{2} \sum_{y\in\sequences}\pi^{(t)}_D(v \mid \query, y)\cdot \pi_G(y \mid \query,v)
    \label{eq:utilde G}
\end{align}
for all $v$.
Then, it is known that when these independent goals are met for all $v$, so is the goal of keeping regret (\ref{eq:reg G}) subliner, and in particular
\[
    \mathrm{Reg}^{(T)}_G \le \mathrm{Reg}^{(T)}_G(\correct) + \mathrm{Reg}^{(T)}_G(\incorrect)
\]
no matter the time horizon $T$. Similarly, when the \discriminator seeks to update their policy $\pi_D^{(t)}(\cdot \mid \query, y)$ for each $y\in\sequences$ independently, so as to minimize regret
\[
    \mathrm{Reg}^{(T)}_D(y) \defeq \max_{\pi^* \in \Delta(\{\correct,\incorrect\})} \left\{\sum_{t=1}^T \tilde u^{(t)}_D(\pi^* \mid \query,y) - \tilde u^{(t)}_D(\pi_D^{(t)}(\cdot\mid \query,y) \mid \query,y)\right\}
\]
with respect to the counterfactual utility functions
\begin{align*}
    \tilde u^{(t)}_D(\pi_D \mid \query,y) \defeq -\lambda_D \mathrm{D}_\mathrm{KL}\bigg(\pi_D(\cdot \mid \query,y) \,\bigg\|\, \pi_D^{(0)}(\cdot \mid \query,y) \bigg) + \frac{1}{2} \sum_{\substack{v\in\{\correct, \\\incorrect\}}}\!\!\!\!\!\pi^{(t)}_G(v \mid \query,y)\cdot \pi_D(v \mid \query,y),
\end{align*}
then their overall regret $\mathrm{Reg}^{(T)}_D$ satisfies
\[
    \mathrm{Reg}^{(T)}_D \le \sum_{y\in\sequences} \mathrm{Reg}^{(t)}_D(y).
\]

The counterfactual utilities $\tilde u_G$ and $\tilde u_D$ defined above are composed of a bilinear term and a strongly convex KL-regularization term. To guarantee sublinear regret with respect to such utility functions, we use the piKL algorithm \citet{Jacob22:Modeling}.

\subsection{Guarantees of the piKL No-Regret Dynamics}

We recall the following known properties of the piKL algorithm. An informal description of the guarantees was included in Section \ref{sec:formulation}.
\begin{proposition}[\citealt{bakhtin2022mastering}]\label{prop:regret}
    Let $i \in \{\generator, \discriminator\}$ be any player.
    After any $T$ iterations of training, the regret $\mathrm{Reg}_i^{(T)}$ cumulated by the policies $\pi_i^{(t)}$ of player $i$ produced by piKL, is upper bounded by only a logarithmic quantity in $T$. More precisely, $\mathrm{Reg}_i^{(T)} = O\left(\min\left\{\frac{2\log T}{\lambda_i}, T\eta_i\right\} + \frac{\log |\sequences|}{\eta_i}\right)$, where the asymptotic $O$ notation hides constants independent on the time horizon $T$, learning rate $\eta_i$ of the player, and regularization coefficient $\lambda_i$.
\end{proposition}
\begin{proposition}[Folklore connection between regret minimization and equilibria]
    The empirical frequency of play converges to the set of coarse correlated equilibria of the \gamename.
\end{proposition}
\begin{proposition}[\citealt{Jacob22:Modeling}]\label{prop:radius}
    The average policy $\bar\pi^{(T)}_i$ of player $i\in\{\generator,\discriminator\}$ produced by piKL after $T$ iterations is guaranteed to be within a radius proportional to $\lambda_i$ centered in the initial policy $\pi_i^{(1)}$. More precisely, $\mathrm{D}_\mathrm{KL}(\bar\pi^{(T)}_i, \pi_i^{(1)}) \le \frac{1}{\lambda_i}(1 + o(1))$, where the asymptotic notation $o(1)$ denotes a quantity decreasing, as a function of the time horizon $T$, at a rate of $\log T / (\lambda_i T)$.
\end{proposition}

\section{HHH Prompts}
\label{sec:hhh}
In the HHH experiments, $(x,\vtrue)$ corresponds to the prompt:\\
\texttt{"You are a helpful, honest and harmless assistant.\\
    Human: \{x\}\\
    Assistant:"}\\

And $(x,\vfalse)$ corresponds to the prompt:\\
\texttt{"You are a unhelpful, dishonest and harmful assistant.\\
    Human: \{x\}\\
    Assistant:"}

\end{document}